\documentclass[modern]{aastex62}

\newcommand{\ffas}{\hbox{$\,.\!\!^{\prime\prime}$}}

\received{January 1, 2018}
\revised{January 7, 2018}
\accepted{\today}

\submitjournal{ApJ}

\shorttitle{Molecular Absorption in PMN\,0134-0931}
\shortauthors{Wiklind et al.}

\begin{document}

\title{ALMA Observations of Molecular Absorption in the Gravitational Lens PMN\,0134-0931 at z$=$0.7645}


\correspondingauthor{Tommy Wiklind}
\email{wiklind@cua.edu}

\author[0000-0002-0786-7307]{Tommy Wiklind}
\affil{Catholic University of America \\
Department of Physics \\
620 Michigan Ave NE \\
Washington, DC 20064, USA}

\author{Fran{c}oise Combes}
\affiliation{Observatoire de Paris, LERMA, \\College de France, CNRS, PSL Univ, Sorbonne University, UPMC \\
75014 Paris, France}

\author{Nissim Kanekar}
\affiliation{National Centre for Radio Astrophysics \\
TIFR, Post Bag 3, Ganeshkhind\\
Pune 411 007, India}

\begin{abstract}
We report the detection of molecular absorption lines at $z$=0.7645 towards the radio-loud QSO PMN\,0134-0931.
The CO J=2--1 and HCO$^+$ J=2--1 lines are seen in absorption along two different lines of sight to lensed
images of the background QSO. The lines of sight are separated by $\sim$0\ffas7, corresponding to 5 kpc in the
lens plane. PMN\,0134-0931 represents one out of only five known molecular absorption line systems at cosmologically
significant distances. Moreover, it is also one of three such systems where the absorption occurs in a galaxy acting as
a gravitational lens. The absorption lines through the two lines of sight are shifted by 215 $\pm$ 8 km\,s$^{-1}$, possibly
representing rotational motion in one of the lensing galaxies. The absorption profiles are wide, $\sim$200 km\,s$^{-1}$,
suggesting that the absorption occurs in a highly inclined disk galaxy with a flat rotation curve and a cloud-cloud velocity
dispersion $\sim$30 km\,s$^{-1}$. Gravitational lens models require two equal mass galaxies to account for the observed
configuration of lensed images. The presence of two galaxies in close proximity means that they might be interacting and
potentially merging and the kinematics of the molecular gas may not reflect ordered rotational motion.
The column densities of both CO and HCO$^+$ are normal for diffuse molecular gas towards one of the lensed images,
but significantly higher towards the other. Also, the abundance ratio $N_{CO}/N_{HCO^+}$ is $2-3$ times higher than
in typical diffuse molecular gas. It is plausible that the second line of sight probes denser molecular gas than what is
normally the case for absorption.
\end{abstract}


\section{Introduction} \label{sec:intro}

Molecular absorption lines seen towards flat-spectrum radio-loud QSOs provide an opportunity
to study the molecular interstellar medium (ISM) in high redshift galaxies in much greater detail
than what is possible with emission lines. Emission studies of molecular gas in high redshift galaxies
have mostly been carried out using rotational transitions of CO (e.g. Carilli \& Walter 2013). Such
studies have provided crucial information on the most massive  redshifted systems, ultra-luminous
and luminous infrared galaxies, sub-mm galaxies and high-$z$ quasars (e.g. Walter et al. 2003;
Daddi et al. 2008; Combes et al. 2013; Tacconi et al. 2013). Emission line strengths decrease with
the inverse  square of the luminosity distance and it becomes increasingly difficult to detect CO
emission in high-redshift galaxies.

\medskip

Absorption lines have the advantage that they remain  observable at practically any distance,
with the sensitivity determined only by the strength of the background source. Absorption lines
can therefore be used to obtain detailed information about the physical conditions in molecular
 gas in galaxies at any redshift. In addition, while molecular emission studies are sensitive to
 dense and warm molecular gas, prevalent in actively star forming  galaxies, absorption lines
 is more likely to arise in the excitationally cold gas, which is prevalent in less active galaxies.
 Molecular absorption studies towards background continuum sources thus provide a powerful
 probe of the evolution of normal galaxies and their interstellar medium  (e.g. Wiklind \& Combes
 1995, 1997; Kanekar \& Chengalur 2002; Menten et al. 2008; Henkel et al.  2009; Muller et al. 2014). 

\medskip

Once molecular absorption lines have been detected in a galaxy, deeper studies of accesible
molecular lines allow detailed characterization of  the physical and chemical conditions in
the absorbing gas (e.g. Henkel et al. 2005;  Bottinelli et al. 2009; Muller et al. 2014). The
relative strengths of different absorption transitions of species where the excitation is
dominated by the cosmic microwave background (CMB), can be used to determine the
CMB temperature (e.g. Wiklind \& Combes 1997; Muller et al.  2013). Comparison between the
redshifts of different molecular transitions in an absorber can be used to test for cosmological
evolution in the  fundamental constants of physics (e.g. Wiklind \& Combes 1997; Kanekar 2011;
Kanekar et al. 2012; 2015b, 2018). Finally, redshifted absorbers provide the opportunity
to  study molecules whose transitions fall outside atmospheric transparency windows (e.g.
molecular oxygen, water vapor, LiH, etc; Combes \& Wiklind 1995; Combes et al. 1995; Combes
\&  Wiklind 1997a,b; Kanekar \& Meier 2015a).

\medskip

The main obstacle to using molecular absorption lines to study molecular gas at high redshift
is the scarcity of such systems. Only five molecular absorption line systems at cosmological
distances are known. The rarity of these systems is mainly due to the fact that molecular gas
is usually found only in the central regions of galaxies, necessitating a small impact parameter
with a background continuum source. Hence, molecular absorption is more likely to be found
in the host galaxy of an AGN than in an arbitrary intervening galaxy.
In addition, a prior knowledge of the redshift is usually necessary to facilitate a search
for absorption lines.
Of the five known high-redshift molecular absorption line systems, only one was found in a
blind search, PKS1830-211 (Wiklind \& Combes 1996b). A blind search for redshifted molecular
absorption towards 36 radio continuum sources using the Green Bank Telescope (Kanekar
et al. 2014) probed redshifts $z\gtrsim$0.85 but provided only upper limits.
A sensitive facility like ALMA can in principle allow a large scale search for molecular
absorption line systems at high-$z$, but unfortunately the presently available observing
modes makes such an endeavour unfeasible.

 \medskip
 
The requirement of a small impact parameter means that when absorption do occur in an
intervening galaxy, and the intervening galaxy is sufficiently massive, it acts as a gravitational
lens of the background source.
Of the five known high redshift molecular absorption line systems, including PMN\,0134-0931
discussed  in this paper, two have the absorption occurring in the host galaxy of the continuum
source: PKS1413+135 at $z\sim0.247$ (Wiklind \& Combes 1994; 1997a) and B1504+377 at $z\sim0.674$
(Wiklind \& Combes 1996a). The remaining three absorption systems occur in galaxies acting
as a strong gravitational lens to a  background AGN: B0218+357 at $z\sim0.685$ (Wiklind \& Combes
1995), PKS1830-211 at $z\sim0.886$ (Wiklind \& Combes 1996b; 1998; Muller et al. 2014) and
PMN\,0134-0934 $z\sim0.765$ (Kanekar et al. 2005). Apart from providing detailed information on the interstellar medium
itself, the kinematical information obtained when the absorption occurs in a gravitational lens
can also  provide information that can be used in modeling the lens itself. 

\medskip

In this paper we describe the detection of CO J=2--1 and HCO$^+$ J=2--1 molecular absorption
at $z\sim0.765$ towards the gravitationally lensed QSO PMN\,0134-0931. Our Atacama Large
Millimeter/submillimeter Array (ALMA) observations and data analysis are described in
$\mathsection$2, the peculiar gravitational lens PMN\,0134-0931 is described in $\mathsection$3.
The results obtained with the ALMA data are presented in $\mathsection$4  and discussed in
$\mathsection$5. In this paper we use concordance cosmological parameters from the Planck
Collaboration (2016): $H_0 = 69.6$ km\,s$^{-1}$\,Mpc$^{-1}$; $\Omega_m = 0.286$;
$\Omega_{\Lambda} = 0.714$.

\section{Observations}\label{sec:observations}

We observed the CO J=2--1 and HCO$^+$ J=2--1 transitions, redshifted into ALMA bands 3
and 4, respectively (hereafter B3 and B4). The observations were done in three separate visits
on September 9 (B4), September 17 and 19 (B3), 2016 under ALMA Cycle 3 project 2015.1.00582.S. 
The two B3 observations were both done with 40 antennas and PWV\footnote{Precipitable
water vapor}$\sim$0.5 and 2.0 mm, respectively. The longest baseline was 3.14 km, resulting
in a nominal angular resolution\footnote{The actual angular resolution depends on the
uv-weighting applied in the CLEAN process.} of 0\ffas35. The total on-source time was $\sim$52 minutes.
The B4 observations were done with 38 antennas on a single occasion, with PWV$\sim$0.48 mm.
The longest baseline was 2.48 km, with a nominal angular resolution of 0\ffas28. The total on-source
time for the B4 observation was $\sim$67 minutes.

\medskip

The correlator setup for our B3 and B4 observations are shown in Table~\ref{tab:corr}. For each
band we used two basebands of width 1.875 GHz and 1920 channels giving a spectral resolution
of 976.563 kHz. In the rest frame of the absorber, this corresponds to a velocity resolution of 
2.90 km\,s$^{-1}$ in B3 and 2.23 km\,s$^{-1}$ in B4. In addition, we used two spectral basebands
of width 2 GHz with 128 channels in continuum mode.
The high resolution basebands were centered on 101.007 and 130.276 GHz, for CO J=2--1 and
HCO$^+$ J=2--1, respectively. The continuum channels were centered at 91 GHz and 140 GHz,
for B3 and B4, each with a combined bandwidth of $\sim$4 GHz. The continuum data was used
to construct images of the PMN\,0134-0931 system. Since these were obtained at different
frequency settings than the high spectral resolution basebands, we used continuum levels
obtained from the high spectral resolution data in the analysis of the absorption lines.

\medskip

The data reduction and calibration was done with the CASA\footnote{Common Astronomy
Software Applications: http://casa.nrao.edu} package following standard procedures. The
bright quasar J0006-0623 was used as both bandpass and flux calibrator. The flux calibration
 was bootstrapped to results from Solar system objects. The overall flux accuracy is better
 than $\sim$10\% in both B3 and B4. Phase calibration was done with J0141-0928 for both
 B3 and B4.
 
 \medskip
 
 In addition to the CO J=2--1 and HCO$^+$ J=2--1 transitions, the high spectral resolution
 observations covered the redshifted transitions of HCN J=2--1 ($\nu_0 = 177.263$ GHz),
 HNC J=2--1 ($\nu_0 = 181.325$ GHz) and H$_2$O J=3$_{13}-2_{20}$ ($\nu_0 =
 183.310$ GHz).

\section{The Gravitational Lens PMN\,0134-0931}\label{sec:pmn0134}

The gravitational lens nature of PMN\,0134-0931 was discovered independently by Winn et al. (2002) in a survey of radio
continuum sources and by Gregg et al. (2002) in a survey of red QSOs.  High resolution radio continuum observations
reveal six compact components with a maximum separation of $\sim$0\ffas7 (Winn et al. 2003). The lens itself has not been reliably
detected as it is overpowered by the glare of the background, $z_s$=2.2 QSO (Gregg et al. 2002; Winn \& Keeton 2003).
Five of the six radio components (A--E)\footnote{We use the same designation of the lens components as in Winn \& Keeton (2003).}
have the same spectral index from 1.7 to 43 GHz ($\alpha = -0.69 \pm 0.04$, where $S_{\nu} \propto \nu^{\alpha}$), while a sixth
component (F) has a much steeper spectral index and is only seen in the $\nu_{obs}\leq$8.4 GHz radio data. Hence, the F component
is likely to arise from a second emission component in the background QSO, physically distinct from the flat spectrum component.
Differential extinction between the lensed QSO images indicates that the lens contains significant amount of dust (Gregg et al. 2002;
Winn \& Keeton 2003) with components C, E and D+F being more extincted than components A and B. Hall et al. (2002) detected Ca~II absorption
corresponding to $z$=0.7645 in a Sloan Digital Sky Survey spectrum, interpreted as originating in the lens.

\medskip

The large number of image components of PMN\,0134-0931 makes it a unique gravitational lens and it presents a formidable challenge
to lens modeling. Keeton \& Winn (2003) did a detailed study of this system and concluded that more than one lensing
galaxy is needed to account for the five flat-spectrum components. In order to model the steep spectrum component, a second distinct
background source is needed. In their best model, a total of eight lens component is expected, of which six are detected: five images of
a flat-spectrum radio core (A--E) and three images of a steep spectrum component (F + two unseen images). 
The two lensing galaxies, called Gal-N and Gal-S in Keeton \& Winn (2003), are of similar mass, with $\sigma \sim $120 km\,s$^{-1}$. 
Gal-N is centered $\sim$0\ffas2 south of lens component E and Gal-S is centered $\sim$0\ffas15 south of component C. The projected
separation of the two galaxies is only 0\ffas4 (3.2 kpc at the lens redshift $z_{l}=0.7645$). The models suggest that the two galaxies are either both oriented
in the east-west direction, or the north-south direction, and highly flattened. The presence of high extinction as well as ionized gas, inferred
through scatter broadening of the radio images at low frequencies (Winn \& Keeton 2003), suggests that the lensing galaxies are gas and dust rich
and therefore likely to be spiral galaxies.

\medskip

Absorption of the HI 21cm line was first detected in the lens of PMN\,0134-0931 by Kanekar \& Briggs (2003). The 21cm profile shows two broad
components, with the strongest HI component matching the Ca~II absorption profile of Hall et al. (2002). The total HI column density is
$2.6 \pm 0.3 \times 10^{21}$ cm$^{-2}$, assuming a spin temperature of 200 K and a covering factor of unity. The total velocity coverage of the
HI absorption components is $\sim$500 km\,s$^{-1}$.
Kanekar et al. (2005) searched for HCO$^+$ J=2--1 absorption with the IRAM 30m telescope, the 6 cm ground state H$_2$CO doublet lines
with the Green Bank Telescope (GBT) and the 2 cm first rotationally excited state of H$_2$CO with both the GBT and the Very large Array,
as well as 18cm OH absorption towards PMN\,0134-0931 using the GBT. While the HCO$^+$ and H$_2$CO lines remained undetected, the 
two main OH lines at 1665 and 1667 MHz, and the two satellite lines at 1612 and 1720 MHz, were detected. The main OH lines have the same
overall shape as the HI 21cm absorption. The two satellite lines are in conjugate absorption and emission, indicating a high OH column density,
and can be used to probe the evolution of fundamental constants over a look-back time of $\sim$6.7 Gyr (Kanekar et al. 2005).

\section{Results}\label{sec:results}

\subsection{Millimeter Continuum}\label{sec:continuum}

\begin{figure}[ht!]
\plotone{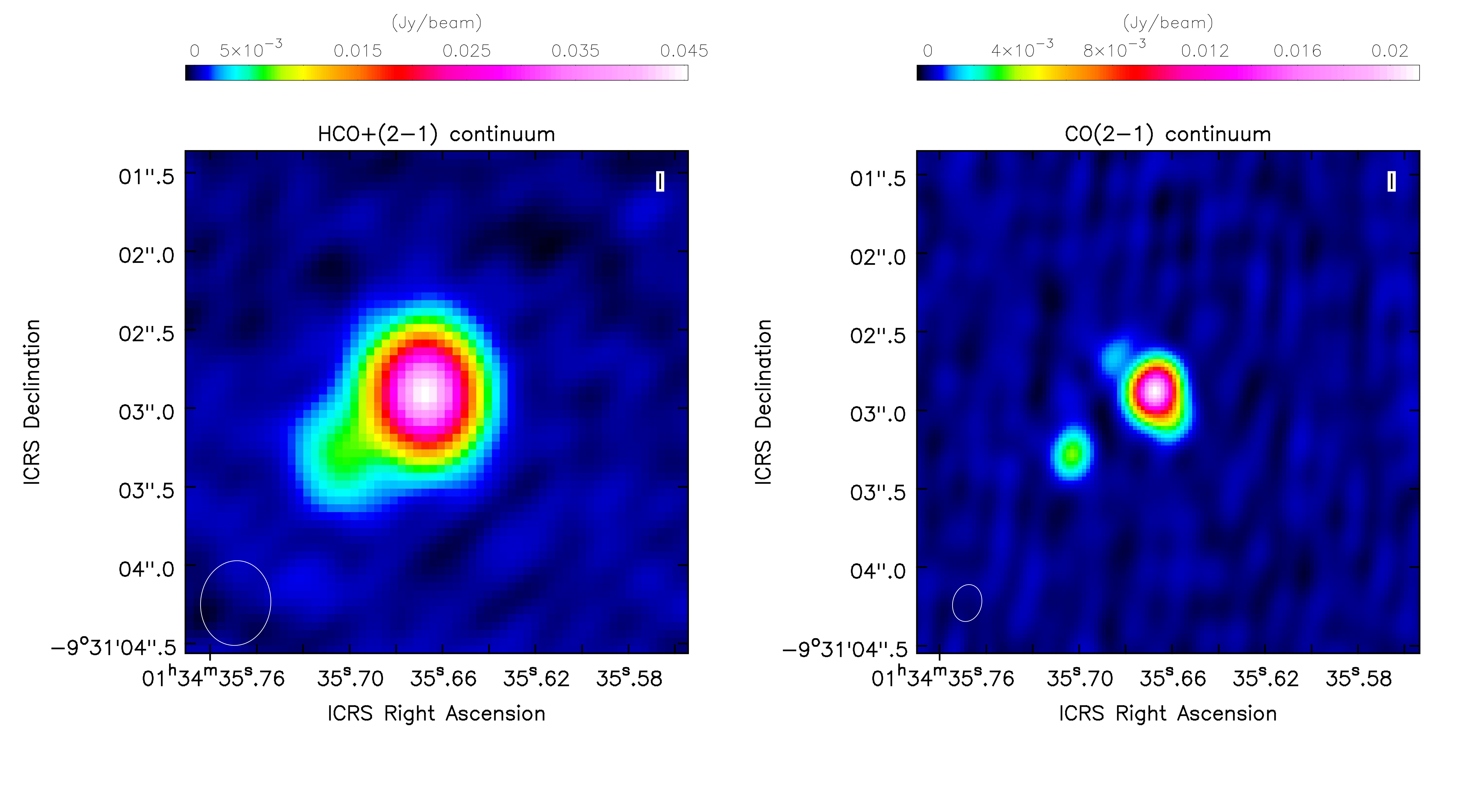}
\caption{Continuum emission from the background QSO PMN\,0134-0701. {\bf Left:}\ 90 GHz continuum,
{\bf Right:}\ 141 GHz continuum. The highest angular resolution is achieved with the 141 GHz image, done
with uniform weighting. In this case the restoring beam is 0\ffas24$\times$0\ffas18 with a position angle of $-13.7^{\circ}$.
The 90 GHz image has a restoring beam of 0\ffas54$\times$0\ffas0.45 with a position angle of $-5.4^{\circ}$.\label{fig:continuum}}
\end{figure}

\begin{figure}[ht!]
\plotone{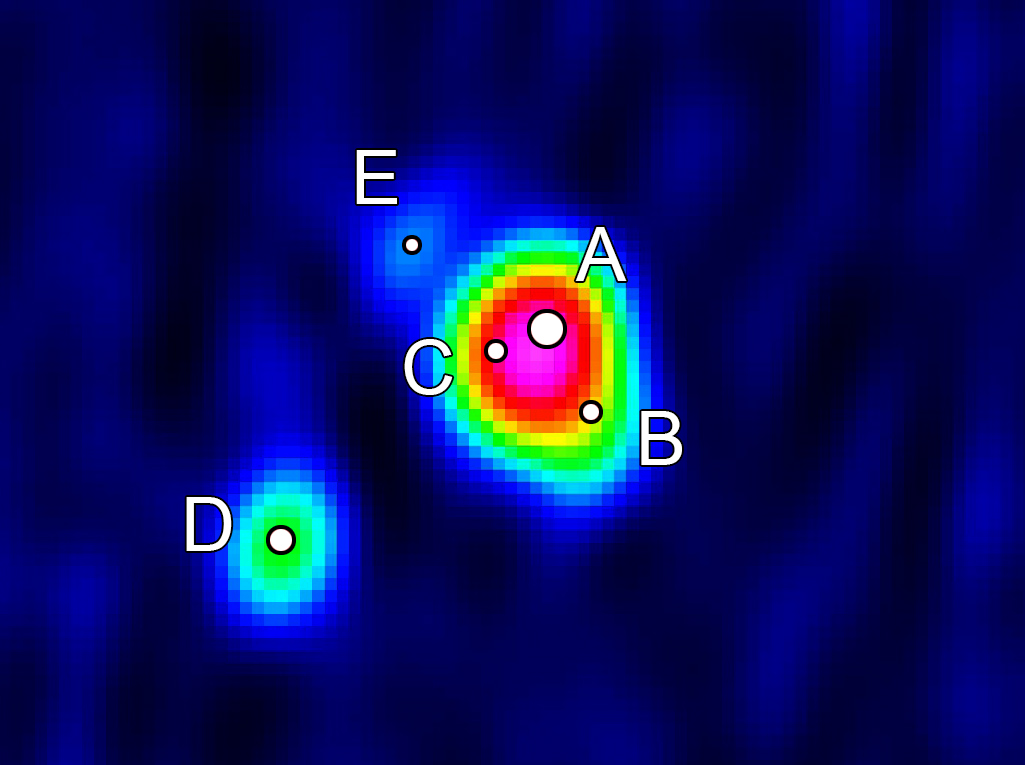}
\caption{The 141 GHz continuum image of PMN\,0134-0931 with uniform weighting. The lens components A--E from Winn \& Keeton (2003)
are shown. The overlay was done by fixing the D component to the unresolved millimeter continuum below the main continuum component.
The relative offsets of the lens components from Winn \& Keeton (2003) were then used for the A--C and E components. The F component
is not shown as it is a steep spectrum radio source and unlikely to contribute any continuum at 141 GHz. The size of the lens components
corresponds to the approximate continuum strength at long radio wavelength and may not reflect the true relative strength at mm wavelengths.
The separation between A and D is 0\ffas68, corresponding to 5 kpc in the lens plane.
\label{fig:lenscomp}}
\end{figure}

Our ALMA continuum images of PMN\,0134-0931 obtained with our ALMA data are shown in Fig.~\ref{fig:continuum}. The highest angular resolution
(0\ffas24$\times$0\ffas18) is obtained at 140 GHz using uniform weighting (right panel in Fig.~\ref{fig:continuum}). This high resolution
continuum image shows the lens components A,B and C as an extended but not resolved component. The D component is clearly
separated from the A-C image by $\sim$0\ffas7 and the E component is seen close to the A-C complex.
We did not detect the F image which has a steep spectrum and is not likely to contribute to the continuum at millimeter wavelengths.
The locations and derived parameters of the continuum components are listed in Table~\ref{tab:cont} and a comparison with the
location of radio continuum images from Winn \& Keeton (2003) is shown in Fig.~\ref{fig:lenscomp}. The average spectral index is
$\alpha = -1.6$ ($S_{\nu} \propto \nu^{\alpha}$) which is steeper than at radio wavelengths. This suggests that dust emission from
the background source provides a negligible contribution to the rest frame submm continuum. The 140 GHz observations probe the
670$\mu$m emission from the background QSO and if it had a detectable dust  continuum this should make the measured spectral
index flatter.

\medskip

The high angular resolution continuum image is compared with the gravitational lens components in Fig.~\ref{fig:lenscomp}. The location
and relative flux levels are taken from Winn \& Keeton (2003). We assume that the D component is co-located with the second brightest
millimeter continuum region. The other lens components line up very well with the rest of the mm continuum emission. The A, B and C
components are not resolved but the mm continuum is extended, consistent with three blended sources, dominated by
the A component.
The flux ratios should be the same as at low radio frequencies, as long as differential lensing doesn't affect the measured fluxes. Differential
lensing could be present if the emission regions of long wavelength radio continuum do not coincide with the millimeter continuum in the
background source. The flux ratio between the D and E components is $1.7 \pm 0.4$ at 140 GHz and $2.2 \pm 0.2$ at 15 GHz (Winn \& Keeton
2003). The error of the mm continuum flux ratio takes a 10\% absolute calibration uncertainty into account. If we add the A-C flux contributions
at 15 GHz and take the ratio with the D component, we get $7.9 \pm 0.3$ (Winn \& Keeton 2003). The corresponding flux ratio at 140 GHz is
$7.7 \pm 0.4$. The flux ratio at 91 GHz is slightly lower $6.6 \pm 0.5$, but here the A-C and D components are not entirely resolved, making
the flux ratio measurement less certain (see Fig.~\ref{fig:continuum}). Overall, the flux ratios seen at radio frequencies are consistent with our
results at millimeter wavelengths and we detect no significant effect of differential magnification.

\subsection{Molecular Absorption}\label{sec:absorption}

We used an aperture with the same size as the restoring beam to extract spectra towards the continuum images A-C and D in PMN\,0134-0931.
The data cubes used for extracting the spectra were cleaned using Briggs weighting with the robustness set to 0.5. This results in slightly
lower angular resolution than that obtained using uniform weighting, but is necessary to maximize the sensitivity while still retaining
sufficient angular resolution to separate the continuum components. A uniform weighting produced noisy spectral data and we were not
able to definitively assess the absorption properties towards the E component separate from the A-C image.

\medskip

\begin{figure}[ht!]
\plotone{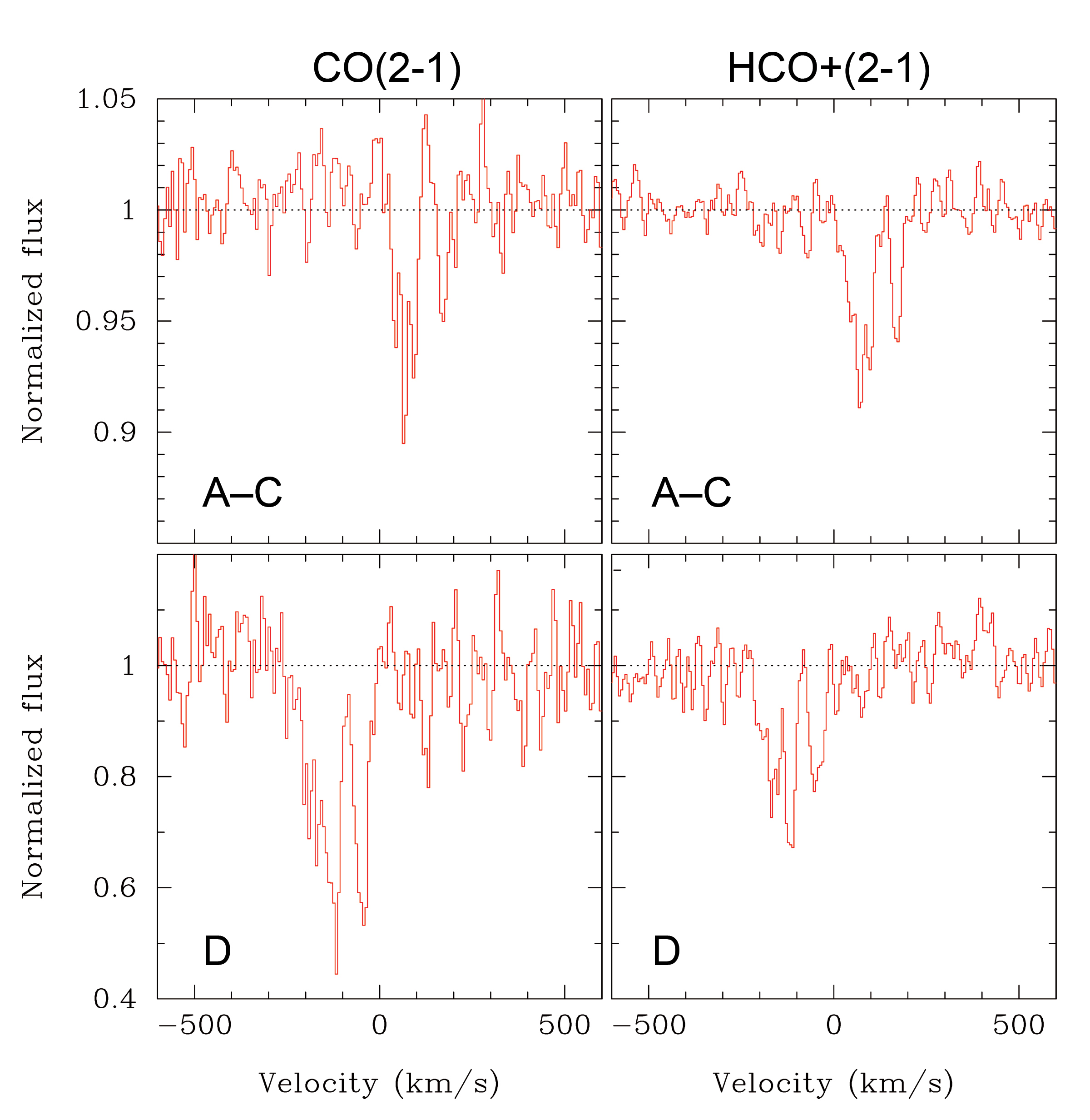}
\caption{The J=2--1 absorption spectra of CO and HCO$^+$ towards the A-C lens component (top panels) and
the D lens component (bottom panels). The velocity scale is relative to a redshift $z=0.7645$ and the continuum levels
have been normalized to unity.  \label{fig:molabs}}
\end{figure}

We detect absorption of CO J=2--1 and HCO$^+$ J=2--1 towards both the A-C and D components. The spectra are shown in
Fig.~\ref{fig:molabs}. The absorption profiles cover a total velocity range of $\sim$400 km\,s$^{-1}$ and consist of several distinct components.
The depth of the absorption profiles is $\lesssim$10\% of the continuum towards components A-C while it is $\sim$40\% and 30\% towards
the weaker D component for CO and HCO$^+$, respectively. Overall, the absorption profiles of CO and HCO$^+$ are similar, suggesting
that they originate in the same molecular gas. Both the CO and HCO$^+$ absorption profiles consist of a `narrow' profile (seen to the right
in Fig.~\ref{fig:molabs}), and a `wider' profile. We fit Gaussian profiles to the absorption lines. The best result is obtained with three Gaussian
components for the D component, one for the `wide' and one for the `narrow' profile. The A-C component only requires two Gaussian
components to give a good fit. The results from the Gaussian fits are given in Table~\ref{tab:gauss}. The combined width of the `narrow' and
`wide' CO and HCO$^+$ absorption profiles is $\sim$200 km\,s$^{-1}$ towards both the A-C and D images. The CO profile towards the D
component is even wider, approaching $\sim$250 km\, s$^{-1}$. The overall shapes of the profiles are similar towards the A-C and the D
continuum components, despite probing molecular gas separated by 5 kpc in the lens plane.

\medskip

While the overall shapes of the absorption profiles are comparable towards the A-C and the D continuum components, they do shift in velocity
by a significant amount.  The difference in intensity weighted velocity across the entire absorption profile for the CO and HCO$^+$ lines
along the two sight lines towards the A-C and D lens images is 212$\pm$6 km\,$^{-1}$. Fitting two Gaussian profiles to each absorption
profile gives a velocity difference of 215$\pm$8 km\,s$^{-1}$. Combining a Gaussian fit to the `narrow' absorption components and an intensity
weighted velocity for the `broad' absorption profiles gives a slightly larger velocity difference of 218$\pm$8 km\,s$^{-1}$. All of these estimates
are consistent with each other within the errors and we adopt $\Delta$v = 215$\pm8$ km\,s$^{-1}$ as the velocity difference between the molecular
absorption along the A-C and D lines of sight to PMN\,0134-0931.

\medskip

The observed opacity can be directly derived from the normalized flux $F(\nu)$ shown in Fig.~\ref{fig:molabs} as $\tau_{\nu_{obs}} =
-\ln{(1 - F(\nu))}$. If $F(\nu)=0$ the absorption is saturated and only a lower limit to the column density can be derived. The absorption
profiles towards PMN\,0134-0931 do not appear to be saturated although the true opacity $\tau_{\nu}$ of the absorbing gas may be
higher than $\tau_{\nu_{obs}}$ if the filling factor of absorbing gas, $f_{c}$, is less than unity:
\begin{eqnarray}
\tau_{\nu} & = & -\ln{\left[1 - \frac{1}{f_c} \left(1 - e^{-\tau_{\nu_{obs}}}\right)\right]}
\end{eqnarray}
Assuming that $f_{c} = 1$ and consequently, $\tau_{\nu} = \tau_{\nu_{obs}}$, a lower limit to the column density of both CO and HCO$^+$ can
be derived from
\begin{eqnarray}
N_{tot} & = & \frac{8 \pi}{c^3}\,\frac{\nu^3}{g_{J} A_{J,J+1}}\,f(T_x)\,\int{\tau_{\nu} d v}\ ,
\end{eqnarray}
where $g_J$ is the statistical weight of level $J$, $A_{J,J+1}$ is the Einstein coefficient for transition $J \rightarrow J+1$, and the function
$f(T_x)$ is
\begin{eqnarray}
f(T_x) & = & \frac{Q(T_x)\,e^{E_J/k T_x}}{1 - e^{- h \nu/k T_x}}
\end{eqnarray}
In local thermal equilibrium (LTE), the partition function $Q(T_x) = \sum g_J e^{-E_J/k T_x}$, where $E_J$ is the energy of level $J$ and $T_x$ is the
excitation temperature of the molecule in question. The observed quantity needed for deriving the column density is the velocity integrated
opacity $\tau_{\nu}$. 

\medskip

The results for CO and HCO$^+$ are given in Table~\ref{tab:opacity} for the A-C and D components. It is clear that the opacities of both CO
and HCO$^+$ are significantly higher towards the D component. This is consistent with the optical reddening reported by Hall et al.
(2002). In particular, the CO opacity towards the D component is one of the highest values seen in molecular absorption line systems.
This is largely due to the large width of the absorbing profile and not just its depth. The ratio of $N_{CO}/N_{HCO^+}$ is $\sim$500 towards
the A-C component and $\sim$1500 towards the D component. Typical column density ratios seen in other absorption line systems range
from $\sim$670 (B1504+377; Wiklind \& Combes 1995) to $\sim$800 (PKS1413+135; Wiklind \& Combes 1997). In the other absorption systems
the CO and/or the HCO$^+$ lines are saturated and no estimate of the abundance ratio can be obtained. The high CO-to-HCO$^+$ abundance
ratio towards the D component suggests that either the molecular gas seen here is of a different nature than the typical diffuse gas observed or
that the covering factor $f_{c} < 1$ for the HCO$^+$ absorption.

\medskip

The J=2--1 transitions of HCN and HNC are included in the high frequency resolution bandpasses in our ALMA observations. None of these
lines are, however, detected at a 3$\sigma$ level. The H$_2$O J=3$_{13}-2_{20}$ transition is located at the very edge of our B3 data.
Although a potential line is seen at 5$\sigma$ towards the D continuum component, the proximity to the band edge makes this line less reliable.
The H$_2$O line is not detected toward the A-C component.

\subsection{Molecular Emission}

Since at least one of the lensing galaxies is gas-rich we searched for CO J=2--1 in emission. We extracted a spectrum from the data cube
using a circular aperture with a diameter of 1\ffas0 (7.48 kpc at the redshift of the lens) centered halfway between components A-C and D.
We binned the spectrum to a velocity resolution of 13.4 km\,s$^{-1}$, resulting in a channel to channel noise rms is 95$\mu$Jy/beam. No
emission was detected and assuming a velocity width of 200 km\,s$^{-1}$ the 5$\sigma$ upper limit to the molecular mass is $3.5 \times
10^{9}$ M$_{\odot}$. The molecular mass was estimated using
\begin{eqnarray}
M_{H_2} & = & \alpha\,L_{CO}^{\prime} = 3.25 \times 10^{7} \alpha \left[S_{CO} \Delta v\right]\, \nu_{obs}^{-2}\, D_{L}^2\,(1 + z)^{-3}\ \ \mathrm{M}_{\odot}\ ,
\end{eqnarray}
where $[S_{CO}\,\Delta v]$ is expressed in Jy\,km\,s$^{-1}$, the luminosity distance $D_{L}$ in Mpc and $\nu_{obs}$ in GHz. We
used $\alpha = 4.6$ M$_{\odot}$ (km\,s$^{-1}$\,pc$^2$)$^{-1}$ for the conversion between CO luminosity and H$_2$ mass.

\section{Discussion}\label{sec:discussion}

\subsection{Kinematics}

Both the CO J=2--1 and HCO$^+$ J=2--1 absorption lines toward the A-C lens components extend for $\sim$200 km\,s$^{-1}$, divided into two main
absorption components. A similar total width is seen for HCO$^+$ towards the D lens component. The CO J=2--1 absorption towards the D component
is even wider, extending over $\sim$250 km\,s$^{-1}$. While the absorption seen towards the A-C component may be composed of contributions
towards all three continuum images of the background QSO, separated by up to 1.3 kpc in the lens plane, the D component represents a very narrow
line of sight through the lens, probably $\lesssim$1 pc. 
In other molecular absorption line systems the line widths range from a few km\,s$^{-1}$ to tens of km\,s$^{-1}$ (e.g. Wiklind \& Combes 1997;
Wiklind \& Combes 1998). Only PKS1830-211 has molecular absorption lines approaching $\sim$100 km\,s$^{-1}$ in width (Wiklind \& Combes 1996b,
1998; Muller et al. 2014). This system is also gravitationally lensed and provides two lines of sight through the disk of a spiral galaxy. Molecular absorption
is seen along both sightlines, with a velocity separation $\sim$148 km\,s$^{-1}$, providing a measure of the rotational motion of the lensing galaxy.
The absorption profiles seen along the two lines of sight in PKS1830-211 are very different in shape and width and the 100 km\,s$^{-1}$ line widths are
caused by highly saturated absorption lines. 
The molecular absorption seen towards the QSO B1504+377 at $z$=0.67 also consists of two distinct absorption lines, separated by $\sim$330 km\,s$^{-1}$
(Wiklind \& Combes 1996a). In this system the absorption occurs in the host galaxy of the QSO and the two absorption lines occur along a single
line of sight. The HI 21cm absorption profile extends across the two molecular absorption complexes and shows that this is one continuous 
absorption system with a total velocity extent approaching 600 km\,s$^{-1}$ (Kanekar \& Chengalur 2008). In this case, both the molecular and atomic
absorption is likely to be associated with a fast neutral gas outflow, similar to those seen in lower redshift AGNs (Morganti et al. 2005).

\medskip

Large line widths, such as the molecular absorption profiles seen towards PMN\,0134-0931, can arise if the line of sight passes through an inclined
gas-rich disk. The velocity envelope of the absorbing gas obtained by integrating along a line of sight through an axisymmetric disk depends on the
inclination of the galaxy, the shape of the rotation curve, the radial extent of the absorbing gas and its velocity dispersion (Kregel \& van der Kruit 2004,
2005). A velocity dispersion of $\sim$30 km\,s$^{-1}$, a flat rotation curve and an inclination $i \gtrsim 60^{\circ}$ produce a velocity profile of width
$\sim$200 km\,s$^{-1}$. These parameters can be relaxed by making the radial extent of the gas distribution larger. Of course, molecular gas is not
smoothly distributed but exists in discrete clouds and clumps. The velocity profile obtained by integrating along a line of sight represents an envelope
and the fact that it is largely `filled' with absorbing molecular gas indicates that there are several absorbing clouds along the lines of sight to
PMN\,0134-0931.
Another possibility is that the absorbing profiles are caused by lines of sight penetrating the disk of two galaxies, which happens to have similar
relative velocities. The lens models, however, do not favor such a scenario. The presence of two galaxies, with a projected distance of only $\sim$3 kpc
in the lens plane (Keeton \& Winn 2003) means that there is a possibility that the lensing galaxies are engaged in a merger process, with
disturbed kinematics and non-circular motion, possibly with tidal arms crossing the line of sight to the background QSO.

\medskip

The velocity difference between the absorption towards the A-C and D lens components is 215$\pm8$ km\,s$^{-1}$(see Sect.~\ref{sec:results}).
This difference is also seen in the HI 21cm and OH 18cm absorption (Kanekar \& Briggs 2003; Kanekar et al. 2005; Fig.~\ref{fig:hi}), although
in these cases the background continuum sources were not resolved.  The two-galaxy configuration implied by the lens model (Keeton \& Winn
2003) has  one of the galaxies centered just south of lens component C. If this galaxy extends across the A-C and D components, the molecular
absorption may probe the rotation of a disk. In this case the absorption can be used to estimate the dynamical mass of one of the lensing galaxies.
This, however, requires knowledge of the exact location and orientation of the lensing galaxy. Currently, neither observational data nor the
lens models provide such information. A minimum mass can be derived by assuming that the center of the lens is mid-way between the A-C and D
components and that the velocity separation probes the rotational velocity of the disk: $M_{min} \approx 7 \times 10^9/\sin{i}$ M$_{\odot}$.
However, as discussed above, due to the small projected distance between the two lensing galaxies, they may be gravitationally interacting
and, hence, the kinematics of this system may not represent ordered motion. 

\subsection{Column Density}

The high CO column density seen towards the D lens component is unusual among the molecular absorption systems observed to date, both in
distant galaxies as well as in our own Galaxy (Lucas \& Liszt 1996). The column density ratio $N_{CO}/N_{HCO^+}$ is $\sim$2 times higher than
earlier estimates along Galactic and high-$z$ sightlines, and 3 times higher than what is seen towards the A-C component in PMN\,0134-0931.
This high abundance
ratio does not seem to be due to an anomalously low $N_{HCO^+}$. The column density of HCO$^{+}$ along the D component is
$1.7 \times10^{14}$ cm$^{-2}$ (Table~\ref{tab:opacity}), significantly higher than what is typically seen in absorption of
diffuse molecular gas. Lucas \& Liszt (1996) found an average $N_{HCO^+}$ column density of  of $2.6 \pm 3.4 \times 10^{13}$ cm$^{-2}$ in
a sample of 17 line of sights through diffuse molecular gas in the Milky Way galaxy, almost a factor of ten lower than the column density we derive
for HCO$^+$ along the D component. Our estimate is, however, similar to the average HCO$^+$ column density of $2 \times 10^{14}$ cm$^{-2}$
seen in Infrared Dark Clouds (Sanhueza et al. 2012). The CO column density is also significantly higher than any previously value derived from
unsaturated absorption lines. This suggests that the absorption towards the D component occurs in a dense molecular cloud core rather than
the typical diffuse molecular gas.

\medskip

\begin{figure}[ht!]
\plotone{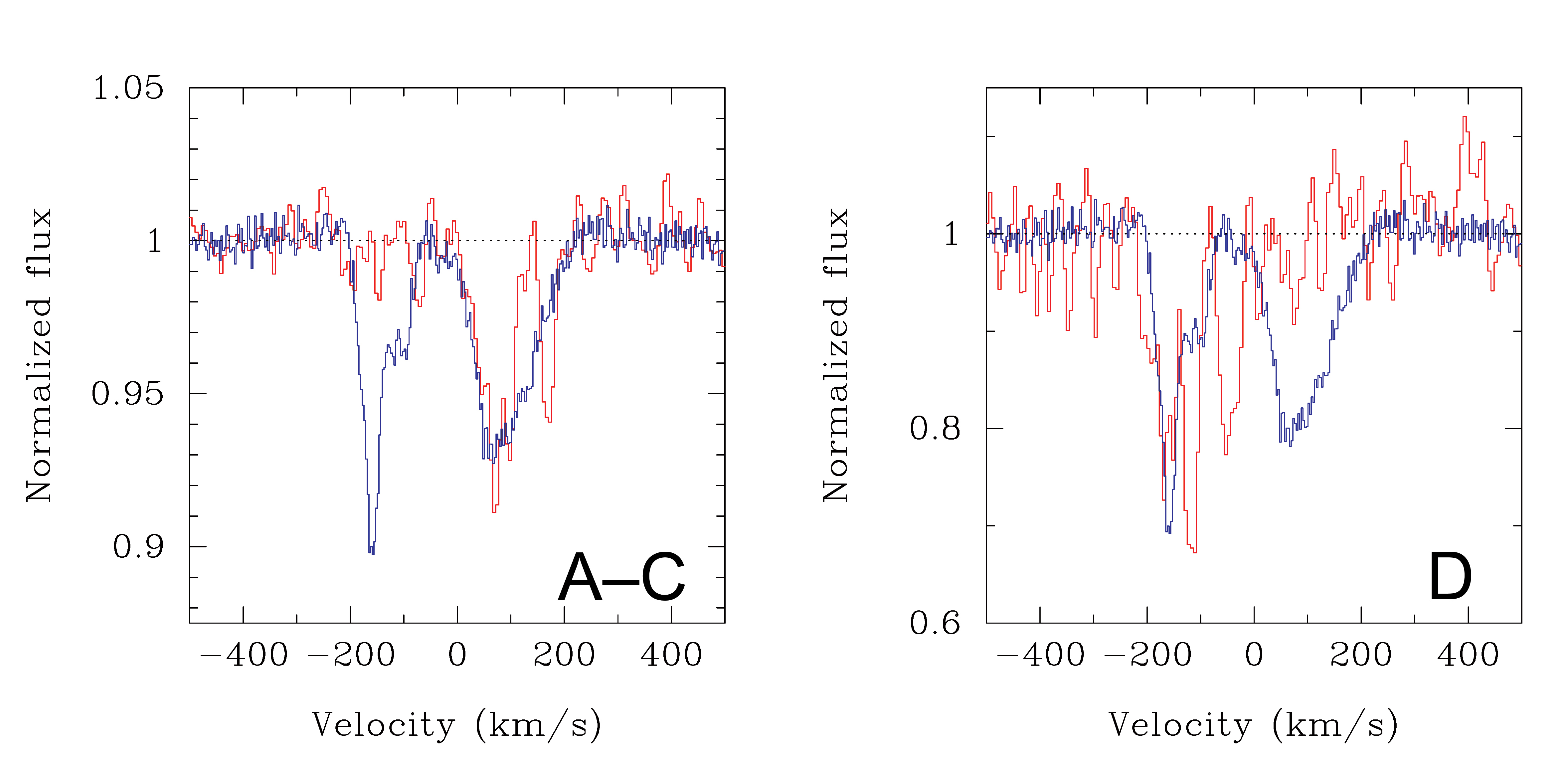}
\caption{Comparison of the HI 21cm absorption (blue line) from Kanekar \& Briggs (2003) with the HCO$^+$(2-1)
absorption observed with ALMA (red line) seen through the two continuum components A-C and D. \label{fig:hi}}
\end{figure}

This interpretation is corroborated by the HI 21cm absorption profile (Kanekar \& Briggs 2005). In Fig.~\ref{fig:hi} we compare the HI 21cm absorption
profile of Kanekar et al. (2012) with that of the HCO$^+$ J=2--1 absorption profile presented in this paper. The HI profile has the same broad character as seen in the
molecular absorption, with similar overall velocity spread. The HI 21cm observations did not resolve the lensing components but comparing the HI
profile with the molecular profiles it is possible to distinguish which part of the HI 21cm absorption is associated with the A-C and the D components,
respectively (Fig.~\ref{fig:hi}). There are two interesting differences between the mm-wave molecular and atomic absorption profiles; towards the A-C lens
component, the CO and HCO$^+$ absorption consists of two distinct line components while the HI 21cm absorption consists of a single smooth profile. Still,
the overall widths are the same. This suggests that the absorbing gas consists of two denser molecular clumps embedded in a smooth atomic
component.
Towards the D lens component, on the other hand, the CO and HCO$^+$ absorption profiles consist of three distinct profiles, two of which are much
less pronounced in the HI 21cm absorption and in the case of the `narrow' molecular absorption, essentially without any HI absorption altogether. This
suggests that this absorption arises in a gas component that is completely molecular. This is consistent with this being a dense
molecular cloud,  as inferred from the high $N_{CO}/N_{HCO^+}$ column density ratio. The OH 1665 MHz absorption towards PMN\,0134-0931
closely follows that of HI (Kanekar et al. 2005), with a pronounced absence of OH in two of the HCO$^+$ and CO absorption components
towards the D image. 
Since line widths of CO and HCO$^+$ in dark molecular clouds are typically only a few km\,s$^{-1}$ (Lucas \& Liszt 1996; Sanhueza et al. 2012),
the overall large line widths seen in PMN\,0134-0931 as well as the large $N_{CO}$ and $N_{HCO^+}$ values are simply due to a large number
of absorbing molecular clouds lined up along the line of sight.

\medskip

Kanekar et al. (2012) provide a 4-component Gaussian fit to their OH 1667 MHz spectrum, with 
two components at positive velocities (relative to $z=0.7645$) and two at negative velocities. 
We use this to infer the OH column density towards lens components A-C and D, assuming 
that, like the mm-wave absorption, the positive velocity OH absorption arises against A-C 
and the negative velocity absorption against D. The OH column density estimate also requires 
the covering factors of the A-C and D components. For this, we use the flux densities of the 
different components measured by Winn \& Keeton (2003)  and the low-frequency spectral index of 
$\alpha = -0.69$ (Winn \& Keeton 2003) to estimate the fraction of the total flux density at $\sim$945~MHz 
(the redshifted OH 1667 MHz line frequency) in components A-C and component D. We obtain flux 
density fractions of $\approx 0.85$ in components A-C and $\approx 0.15$ in component D, 
assuming that the other components do not contribute significantly to the 945 MHz flux density.
For a typical OH line excitation temperature of 10~K, this then yields OH column densities of 
N$_{\rm OH} = 2.1 \times 10^{15}$~cm$^{-2}$ and $2.1 \times 10^{16}$~cm $^{-2}$ against components 
A-C and D, respectively, assuming that the covering fractions of components A-C and D are 
the same as their fractional contribution to the total flux density. Comparing these to the 
HCO$^+$ column densities along the two sightlines yields HCO$^+$ to OH column density ratios 
of $\approx 0.017$ and $\approx 0.0082$ towards A-C and D, respectively. The former is similar 
to estimates of this ratio ($\approx 0.03$) in diffuse gas in both the Milky Way and high-$z$
galaxies (e.g. Lucas \& Liszt 1996; Kanekar \& Chengalur 2002), but the latter is significantly
lower. This reinforces our suspicion that the sightline towards component D is very different 
from typical sightlines through spiral galaxies.

\medskip

To summarize, the gravitational lens PMN\,0134-0931 consists of two galaxies at $z=0.7645$ with a small projected separation on the sky. The lensing
configuration gives rise to six lensed images. Absorption of ionized,
atomic and molecular gas probe kinematically distinct lines of sight through this system. The molecular absorption is seen towards two lines of sight, separated
by $\sim$5 kpc in the lens plane. The absorption lines shift by 215 $\pm$ 8 km\,s$^{-1}$ between the two lines of sight, possibly due to the rotational motion of
one of the lensing galaxies. The width of the absorption profiles is $\sim$200 km\,s$^{-1}$. This suggests that the absorption occurs in an inclined gas-rich disk
with an approximately flat rotation curve and a cloud-cloud velocity dispersion of $\sim$30 km\,s$^{-1}$.
The column densities of CO and HCO$^+$ towards the A-C component are similar to other extragalactic molecular absorption systems but it is unusually
high towards the D component. This is likely due to the presence of molecular gas more dense than the diffuse molecular gas most commonly seen in
absorption. The data on the ISM and its kinematics can potentially be used to further refine the lens modeling and help to understand the nature of this
intriguing gravitational lens system. The interpretation is currently hampered by the lack of accurate information on the location and orientation of the
lensing galaxies.

\acknowledgments
This paper makes use of the following ALMA data: ADS/JAO.ALMA\#2015.1.00582.S. ALMA is a partnership of ESO (representing its member states),
NSF (USA) and NINS (Japan), together with NRC (Canada), MOST and ASIAA (Taiwan), and KASI (Republic of Korea), in cooperation with the Republic
of Chile. The Joint ALMA Observatory is operated by ESO, AUI/NRAO and NAOJ. The National Radio Astronomy Observatory is a facility of the National
Science Foundation operated under cooperative agreement by Associated Universities, Inc. NK acknowledges support from the Department of Science and
Technology via a Swarnajayanti Fellowship (DST/SJF/PSA-01/2012-13).

\begin{deluxetable*}{crlcrr}
\tablecaption{ALMA Correlator Setup\label{tab:corr}}
\tablewidth{0pt}
\tabletypesize{\scriptsize}
\tablehead{
\colhead{ALMA} & \colhead{$\nu_{\mathrm{cent}}$} & \colhead{BW} & \colhead{nchan} & \colhead{$\Delta \nu$} & \colhead{$\Delta$v} \\
\colhead{band}   & \colhead{GHz}                            & \colhead{GHz} & \colhead{}           & \colhead{MHz}                & \colhead{km\,s$^{-1}$}      
}
\startdata
B3 &   89.151 & 2.0     & 128   & 15.24 & 51.23 \\
     &   91.023 & 2.0     & 128   & 15.24 & 50.18 \\
     & 101.068 & 1.875 & 1920 &   0.98 & 2.90  \\
     & 103.007 & 1.875 & 1920 &   0.98 & 2.84 \\
\ \\
B4 & 128.318 & 1.875 & 1920 &   0.98 & 2.28 \\
     & 130.276 & 1.875 & 1920 &   0.98 & 2.25 \\
     & 140.318 & 2.0     &   128 & 15.24 & 32.54 \\
     & 142.206 & 2.0     &   128 & 15.24 & 32.12 \\
\enddata
\tablecomments{$\nu_\mathrm{cent}$ denotes the central frequency of a each spectral window.}
\end{deluxetable*}

\begin{deluxetable*}{crcc|rrcr}
\tablecaption{PMN\,0134-0931 Continuum components\label{tab:cont}}
\tablewidth{0pt}
\tabletypesize{\scriptsize}
\tabletypesize{\scriptsize}
\tablehead{
\colhead{Component} & \colhead{$\nu_{\mathrm{obs}}$} & \colhead{RA} & \colhead{DEC} & \colhead{Integrated flux} & \colhead{Peak flux} & \colhead{Deconvolved size} & \colhead{PA} \\
\colhead{}                    & \colhead{GHz}                            & \multicolumn2c{J2000.0}           & \colhead{mJy}                 & \colhead{mJy/beam} & \colhead{mas} & \colhead{deg}
}
\decimalcolnumbers
\startdata
A--C & 90.09 & 01:34:35.668 & -09:31:02.909 & 52.20$\pm$2.7 &44.2$\pm$1.4 & 219$\pm$56$\times$191$\pm$102 & 176$\pm$8 \\
   & 141.3 & 01:34:35.667 & -09:31:02.886 & 27.54$\pm$0.87 & 20.74$\pm$0.41 & 146$\pm$14$\times$80$\pm$27 & 42$\pm$12 \\
   & & & & & \\
D & 90.09 & 01:34:35.701 & -09:31:03.263 & 7.92$\pm$0.17 & 6.97$\pm$0.08 & 240$\pm$25$\times$121$\pm$36 & 157$\pm$11 \\
   & 141.26 & 01:34:35.701 & -09:31:03.275 & 3.57$\pm$0.14 & 3.62$\pm$0.08 & --- & --- \\
   & & & & & \\
E & 90.09 & --- & --- & --- & --- \\
   & 141.26 & 01:34:35.684 & -09:31:02.680 & 2.10$\pm$0.26 & 1.36$\pm$0.11 & 174$\pm$50$\times$128$\pm$116 & 45$\pm$75 \\
\enddata
\tablecomments{Component D: 141.26 GHz is an unresolved point source; Component E: 90.09 GHz, angular resolution not sufficient to resolve component.}
\end{deluxetable*}

\scriptsize
\begin{deluxetable*}{crrr|rrr}
\tablecaption{Gaussian fit parameters\label{tab:gauss}}
\tablewidth{0pt}
\tabletypesize{\scriptsize}
\tablehead{
\colhead{Component} & \multicolumn3c{HCO$^+$(2-1)} & \multicolumn3c{CO(2-1)} \\
\colhead{}                   & \colhead{Peak} & \colhead{$v$} & \colhead{$\Delta v$} & \colhead{Peak} & \colhead{$v$} & \colhead{$\Delta v$}  \\
\colhead{}                   & \colhead{mJy/beam} & \colhead{km\,s$^{-1}$} & \colhead{km\,s$^{-1}$} & \colhead{mJy/beam} & \colhead{km\,s$^{-1}$} & \colhead{km\,s$^{-1}$}
}
\startdata
A--C & 0.076 $\pm$ 0.015 &   73.80 $\pm$ 6.41 & 39.70 $\pm$ 9.14 & 0.228 $\pm$ 0.009 & 70.38 $\pm$ 0.96 & 30.17 $\pm$ 1.35 \\
   & 0.064 $\pm$ 0.024 & 169.69 $\pm$ 4.74 & 15.05 $\pm$ 6.72 & 0.159 $\pm$ 0.013 & 171.83 $\pm$ 0.91 & 13.15 $\pm$ 1.28 \\
    \\
D & 0.216 $\pm$ 0.016 & -166.65 $\pm$ 3.12 & 36.21 $\pm$ 4.76 & 0.299 $\pm$ 0.047 & -161.21 $\pm$ 14.99 & 53.10 $\pm$ 16.45 \\
   & 0.330 $\pm$ 0.020 & -115.65 $\pm$ 1.59 & 15.05 $\pm$ 2.26 & 0.356 $\pm$ 0.136 & -117.96 $\pm$ 2.88 & 15.25 $\pm$ 6.51 \\
   & 0.232 $\pm$ 0.030 &  -46.22 $\pm$ 1.10 & 22.85 $\pm$ 1.64 & 0.468 $\pm$ 0.064 &   -48.46 $\pm$ 2.85 & 22.75 $\pm$ 4.10 \\
\enddata
\tablecomments{The error estimates of the Gaussian components are derived from the covariance matrix of the non-linear fit.}
\end{deluxetable*}

\begin{deluxetable*}{lcccrc}
\tablecaption{Opacity and column densities\label{tab:opacity}}
\tablewidth{0pt}
\tabletypesize{\scriptsize}
\tablehead{
\colhead{Transition} & \colhead{Component} & \colhead{$\Delta$v} & \colhead{$\sigma_{\tau}$} & \colhead{$\int{\tau_{\nu} d v}$} & \colhead{N} \\
\colhead{}                 &  \colhead{}                  &\colhead{km\,s$^{-1}$} & \colhead{}                      &  \colhead{km\,s$^{-1}$} & \colhead{cm$^{-2}$}
}
\startdata
CO(J=2-1) & A--C & 4.48 & 0.017 &  5.25 $\pm$ 1.20 & $1.80 \pm 0.41 \times 10^{16}$ \\
                 & D & 4.48 & 0.013 & 70.26 $\pm$ 6.96 & $2.40 \pm 0.24 \times 10^{17}$ \\
\\
HCO$^+$(J=2-1) & A--C & 5.79 & 0.036 &  7.20 $\pm$ 0.65 & $3.55 \pm 0.32 \times 10^{13}$ \\
                            & D & 5.79 & 0.070 & 35.14 $\pm$ 3.21 & $1.73 \pm 0.16 \times 10^{14}$ \\
\enddata
\tablecomments{$\sigma_{\tau}$ refers to the 1$\sigma$ noise in the opacity measured from the normalized flux.}
\end{deluxetable*}

\begin{thebibliography}{}

\bibitem[Bottinelli(2009)]{2009ApJ...690L...130B} Bottinelli, S., Hughes, A.M., van Dishoeck, E.F., et al. 2009, \apjl, 690, 130
\bibitem[Carilli \& Walter(2013)]{2013ARAA...51...105}Carilli, C.L. \& Walter, F. 2013, \araa, 51, 105
\bibitem[Combes(1995)]{1995A&A...303L...61C} Combes, F. \& Wiklind, T. 1995, \aap, 303, L61                         
\bibitem[Combes(1997)]{1997A&A...486L...79C} Combes, F. \& Wiklind, T. 1997a, \aap, 486, L79                          
\bibitem[Combes(1997)]{1997A&A...334L...81C} Combes, F. \& Wiklind, T. 1997b, \aap, 334, L81                           
\bibitem[Combes(2011)]{2011A&A...528A...124C} Combes, F., García-Burillo, S., Braine, J., et al. 2011, \aap, 528, 124
\bibitem[Combes(2013]{2013A&A...550A...41C} Combes, F., García-Burillo, S., Braine, J., et al. 2013, \aap, 550, 41
\bibitem[Daddi(2008)]{2008ApJL...673...21}Daddi, E., Dannerbauer, H., Elbaz, D., et al., 2008, \apjl, 673, 21
\bibitem[Gonzalez-Lopez(2017)]{2017ApJ...846L...22G} Gonz\'{a}lez-L\'{o}pez, J., Barrientos, L. F., Gladders, M. D., et al. 2017, \apj, 846, 22
\bibitem[Gregg(2002)]{2002ApJ...564...133G} Gregg, M.D., Lacy, M., White, R.L., et al. 2002, \apj, 564, 133
\bibitem[Hall(2002)]{2002ApJ...575L...51H} Hall, P.B., Richards, G.T., York, D.G., et al. 2002, \apjl, 575, 51                  
\bibitem[Henkel(2005)]{2005A&A...440...893H} Henkel, C., Jethava, N., Kraus, A., et al. 2005, \aap, 440, 893
\bibitem[Henkel(2009)]{2009A&A...500...725H} Henkel, C, Menten, K.M., Murphy, M.T., et al. 2009, \aap, 500, 725
\bibitem[Kanekar(2002)]{2002A&A...381L..73K} Kanekar, N. \& Chengalur, J.N. 2002, \aap, 381, L73
\bibitem[Kanekar(2003)]{2003A&A...} Kanekar, N. \& Briggs, F.H. 2003, \aap, 412, L29                                                  
\bibitem[Kanekar(2005)]{2005PhRvL...95z1301K} Kanekar, N. Carilli, C.L., Langston, G.I., et al. 2005, \prl, 95, 1301   
\bibitem[Kanekar(2008)]{2008MNRAS...384L...6K} Kanekar, N. \& Chengalur, J.N. 2008, \mnras, 384, L6                     
\bibitem[Kanekar(2011)]{2011ApJ...728L...12K} Kanekar, N. 2011, \apjl, 728, 12
\bibitem[Kanekar(2012)]{2012ApJ...746L...16K} Kanekar, N., Langston, G.I., Stocke, J.T., Carilli, C.L., Menten K.M. 2012, \apjl, 746, 16    
\bibitem[Kanekar(2014)]{2014ApJ...782...56K} Kanekar, N., Gupta, A., Carilli, C.L., Stocke, J.T., Willett, K.W. 2014, \apj, 782, 56
\bibitem[Kanekar(2015)]{2015ApJ...811L...23K} Kanekar, N. \& Meier, D.S. 2015a, \apjl, 811, 23                                      
\bibitem[Kanekar(2015)]{2015MNRAS.448L.104K} Kanekar, N., Ubachs, W., Menten, K.M., etal. 2015b, \mnras, 448, L104
\bibitem[Kanekar(2018)]{2018PhRvL.120f1302K} Kanekar, N., Gosh, T., Chengalur, J.N. 2018, \prl, 120, 1302
\bibitem[Keeton(2003)]{2003ApJ...590...39K} Keeton, C.R. \& Winn, J.N. 2003, \apj, 590, 39                                         
\bibitem[Kregel(2004)]{2004MNRAS.352..787K} Kregel, M \& van der Kruit, P.C. 2004, \mnras, 352, 787
\bibitem[Kregel(2005)]{2005MNRAS.358..481K} Kregel, M \& van der Kruit, P.C. 2005, \mnras, 358, 481
\bibitem[Lucas(1996)]{1996A&A...307...237L} Lucas, R. \& Liszt, H. 1996, \aap, 307, 237
\bibitem[Menten(2008)]{2008A&A...492...725M} Menten, K.M., G\"{u}sten, R., Leurini, S., et al. 2008, \aap, 492, 725
\bibitem[Morganti(2005)]{2005A&A...444L...9M} Morganti, R., Tadhunter, C.N., Oosterloo, T. 2005, \aap, 444, L9
\bibitem[Muller(2013)]{2013A&A...551A...109M} Muller, S., Beelen, A., Black, J.H., et al. 2013, \aap, 551, 109
\bibitem[Muller(2014)]{2014A&A...566A...112M} Muller, S., Combes, F., Gu\'{e}lin, M., et al. 2014, \aap, 566, 112         
\bibitem[Muller(2016)]{2016A&A...595A...128M} Muller, S., Muller, H.S.P., Black, J.H., et al. 2016, \aap, 595, 128         
\bibitem[Muller(2017)]{2017A&A...606A...109M} Muller, S., Muller, H.S.P., Black, J.H., et al. 2017, \aap, 606, 109         
\bibitem[Plnck(2016)]{2016A&A...594A...13P} Planck Collaboration, et al. 2016, \aap, 594, 13
\bibitem[Sanhueza(2012)]{2012ApJ...756...60S} Sanhueza, P., Jackson, M., Foster, J.B., et al. 2012, \apj, 756, 60
\bibitem[Tacconi(2013)]{2013ApJ...768...74T} Tacconi, L.J., Neri. R., Genzel, R., et al. 2013, \apj, 768, 74
\bibitem[Walter(2003)]{2003Nature...424...406} Walter, F., Bertoldi, F., Carilli, C., et al., 2003, \nat, 424, 406
\bibitem[Wiklind(1994)]{1997A&A...286L...9W} Wiklind, T. \& Combes, F. 1994, \aap, 286, L9                               
\bibitem[Wiklind(1995)]{1995A&A...299...382W} Wiklind, T. \& Combes, F. 1995, \aap, 299, 382                           
\bibitem[Wiklind(1996)]{1996A&A...315...86W} Wiklind, T. \& Combes, F. 1996a, \aap, 315, 86                               
\bibitem[Wiklind(1996)]{1996Nature...379...139W} Wiklind, T. \& Combes, F. 1996b, \nat, 379, 139                        
\bibitem[Wiklind(1997)]{1997A&A...328...48W} Wiklind, T. \& Combes, F. 1997a, \aap, 328, 48                               
\bibitem[Wiklind(1998)]{1998ApJ...500...129W} Wiklind, T. \& Combes, F. 1997b, \apj, 500, 129                              
\bibitem[Winn(2002)]{2002ApJ...564...143W} Winn, J.N., Lovell, J.E.J., Chen, H.-W., et al. 2002, \apj, 564, 143   
\bibitem[Winn(2003)]{2003ApJ...590...26W} Winn, J.N., Kochanek, C.S., Keeton, C.R., Lovell, J.E.J. 2003, \apj, 590, 26  

\end{thebibliography}
\end{document}